# A New Similairty Measure For Spatial Personalization

Saida Aissa[1] and Mohamed Salah Gouider[2]

[1]BESTMOD Laboratory, High Institute of Management, Tunisia
`saida.aissi@yahoo.fr`
[2]BESTMOD Laboratory, High Institute of Management, Tunisia
`ms.gouider@yahoo.fr`

## ABSTRACT

*Extracting the relevant information by exploiting the spatial data warehouse becomes increasingly hard. In fact, because of the enormous amount of data stored in the spatial data warehouse, the user, usually, don't know what part of the cube contain the relevant information and what the forthcoming query should be.*
*As a solution, we propose to study the similarity between the behaviors of the users, in term of the spatial MDX queries launched on the system, as a basis to recommend the next relevant MDX query to the current user.*
*This paper introduces a new similarity measure for comparing spatial MDX queries. The proposed similarity measure could directly support the development of spatial personalization approaches. The proposed similarity measure takes into account the basic components of the similarity assessment models: the topology, the direction and the distance*

## 1. INTRODUCTION

SOLAP users exploit data cube by launching a sequence of queries over the spatial data warehouse. The process of handling data cube became very tedious and hard because the user usually have no idea of what part of the cube contain the information and what the forthcoming query should be [1]. Applying recommendation technologies is of particular relevance to the domain of spatial multidimensional databases since the SOLAP analysis became a hard task for the users.

In OLAP systems, recommandation is defined as the process that proposes a new OLAP query to the user according to his preferences and needs in order to facilitate the analysis process and assist the user during the exploration of the OLAP system. Giacometti et al [2, 3] and Jerbi et al [4, 5] define recommendation as a process that exploits user's previous queries on the cube and what they did during the previous session in order to recommend the next query to the actual user. In recommendation process, the recommended query is different from the initial query due to different user interests.

Recommendation approaches are classified into two main categories: Collaborative recommendation approaches based on query log analyzes [2, 3, 6, 7] and Individual recommendation approaches based on user profile analyzes: the system provides alternatives and anticipated recommended query taking into account the user context [4, 5].

SOLAP users have specific needs, preferences and goals. However, SOLAP personalization is a search field not well exploited. In fact, the only work proposing personalization of SOLAP systems is proposed by Glorio et al. [8, 9] who present an adaptation of the DW schema by integrating the required spatiality at the conceptual level.

To better support this process, we propose to use a collaborative approach for recommending SOLAP queries. The idea is to exploit the previous operations executed by the other users during their former navigations on the cube, and to use this information to recommend to the current user a relevant anticipated query.

The objectif of this paper is not to present the complete approach which is under development, but to present a new similarity measure between the spatial MDX queries. This similarity measure will be included in our recommendation approach for spatial personalization. To the best of our knowledge, our work is the first work proposing a similarity measure between spatial MDX queries for spatial personalization.

The remainder of this paper is organized as follows: Section 2 presents the basic definitions in the context of spatial data warehouses and SOLAP systems. Section 3 presents the different spatial similarity assessment models proposed in the literature. Section 4 presents the proposal of the new similarity measure, section 5 presents the performance evaluation and section 6 concludes the paper.

## 2. BASIC DEFINITIONS (CUBE, QUERY REFERENCES)

In this paper, we present an approach for computing the similarity between spatial MDX queries. We first begin by presenting the basic definitions, then, we present the different spatial similarity assessment models in the literature and finally, we present our proposal of the new similarity measure for spatial personalization.

*Cube*
A fact is an instance of the fact table. A fact table is called F. A fact table is an instance of a relation. With $F = (N_{01},......., N_{0N}, m_1,...m_N)$ Or, $(m_1,...m_N)$ is a set of numeric attributes representing the different values and measures, $(N_{01},......., N_{0N})$ have the primary key of F, where For each $i \in [1,N]$, Ni is the primary key of the dimension *Di*.
A fact is spatial if it represents a spatial join between two or more spatial dimensions.

**Example:** We present in Figure 1 a star schema allowing to the analysis of the crop (production) by region, by period and by product (type of harvesting). The star schema diagram is presented using the formalism of [10].

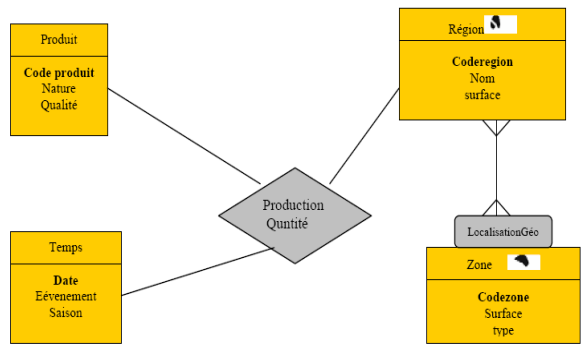

Figure 1. A star schema for the analysis of the production

The fact table for the analysis of the quantity of the production according to the dimensions area and time is presented as follows: *F=(codeproduct, Date, coderegion, Quantity)*

*Query references*
Given a cube C and an MDX query *qc* over C. We define the set of the references corresponding to an MDX query as follows:

$Rqc = \{R_1, R_2, R_3 \ldots R_n, M_1, M_2, \ldots M_N\}$ with:

$\{R_1, R_2, R_3 \ldots R_n\}$: is the set of dimension members *Di* invoked in the MDX query. It represents the set of members of the dimension *Di* that is deduced from the SELECT and WHERE clause.
$\{M_1, M_2, \ldots M_N\}$: is the set of measures extracted from the MDX query

**Example:** Given the following MDX queries:

*q1*: SELECT {[Product]. [All Products]. [No biological]} ON COLUMNS,
FROM [Production]
WHERE {[Measures]. [Quantity]} AND {[Region]. [All Region]. [Region1]. [Zone1]

The references of the query *qc* are *Rq1*= {no biological, Zone 1, all time, quantity}

*q2*: SELECT {[Product]. [All Products]. [Biological]} ON COLUMNS,
FROM [Production]
WHERE {[Measures]. [Quantity]} AND {[Time]. [All Time]. [2012] intersect} [Zone 3]}

The references of the query *q2* are *Rq2* = {biological, All Region, 2012, amount}

*q3*: SELECT {[Product]. [All Products]. [Biological]} ON COLUMNS,
FROM [Production]
WHERE {[Measures]. [Quantity]} AND {[Region]. [All Region]. [Region1]. [Zone 2] intersect [Zone 3]}

The references of the query *q3* are *Rq3*: {biological, All Region, Zone 2, Zone 3, quantity, intersect}

## 3. SAPTIAL SIMAILRITY ASSESSMENT APPROAHCES

Two main approaches are adopted in the spatial similarity assessment: the conceptual neighborhood approach and the projection-based approach [11].

In the conceptual neighborhood approach, the similarity between two concepts is measured according to the distance between these two concepts in a network. The approach computes the shortest path between two concepts in the network basing on a transformation model. The similarity value depends on the number of edges between the concepts [12, 13].

[14] propose changes on the topological relationship based on the Egenhofer's 9-intersection model. They presented a conceptual neighborhood graph of the topological relationship. Possible changes are presented as a sequence of movements over the neighborhood network. If the distance from disjoin (x, y) to meet (x, y) is set as 1, the distance from disjoin (x, y) to covers (x, y) should be 3.

[15] propose a conceptual neighborhood network of 169 possible spatial relations between rectangles. [16] capture the spatial relationship similarity presented in the Chang and Lee's graph by combining the conceptual neighborhood model. They describe the similarity measuring process as follows: "one scene is transformed into another through a sequence of gradual changes of spatial relations". The number of changes required yields a measure that is compared against others, or against a pre-existing scale. Two scenes that require a large number of changes are less similar than scenes that require fewer changes."

The projection-based models divide the space into four main directions (north, west, south, and east) and into four secondary directions (northwest, southwest, southeast, and northeast).
The projection-based approaches project spatial objects and their relations onto a vector space or a matrix space. This changes the problem of similarity assessment from the comparison of objects in spatial scenes to the comparison of the vector or matrix space. In fact, Chang defines three types of similarity values, value-0, value-1 and value-2. value-0 is obtained when two objects have the same relationship on either the x axis or the y-axis. However, for value-1, the two compared objects should have the same relations on both the x-axis and y-axis. Type-2 needs that the two compared objects have the same relations as well as they the same rank of the relative positions.

For distance similarity measurement, [17] propose to combine the projection-based approach and the conceptual neighborhood approach. They propose to use a 3*3 matrix presenting the projection of the nine directions (north, northwest, west, southwest, south, southeast, east, northeast, and same). The projection of an object on the matrix reflects the position of the object regarding the nine directions.
The similarity is computed using the least cost of transformation of one direction-relation matrix into another.

[11] propose the TDD model (Topology-Direction-Distance). The proposed model takes into account both the commonality and the difference to compute the similarity assessment and it affects an order of priority on the topology, the direction and the distance (TDD) into the spatial similarity assessment.

## 4. SIMILARITY MEASURE FOR COMPARING SPATIAL MDX QUERIES

Three main types of spatial relations between objects are defined in the literature: topological relations, metric relations and directional relations (which express a direction: North, West ...)[11, 17].
Thus, measuring the spatial distance between two spatial queries, return to measuring the topological distance, the metric distance and the directional distance between the spatial objects invoked in each query.

### 4.1. Topological distance between spatial MDX queries

#### 4.1.1. Spatial scene

A spatial scene is a spatial representation of one or more spatial objects and the topological relationship between them [12]. A spatial object can be a point, a line or a polygon.

**Definition**

Given a spatial scene $SS_{ab}$ invoked by a query $qc$. $SS_{ab}$ is modeled as follows:

$$SS_{ab}= (Ob_a, Rt_{ab}, Ob_b)$$

With:

$Ob_a$, $Ob_b$ are spatial objects $Ob_a, Ob_b \in$ {point, ligne, polygone}
$Rt_{ab}$ is a topological relation with $Rt_{ab} \in$ {meet, covers,……….}

An MDX query launched by a user can invoke one or more spatial scenes at the clause (WHERE) of the query. Thus we model the spatial scenes invoked by a query as follows:

$$SSqc=\{SS1, SS2,…SSn\}$$

**Example:** The following figure presents an example of three spatial scenes

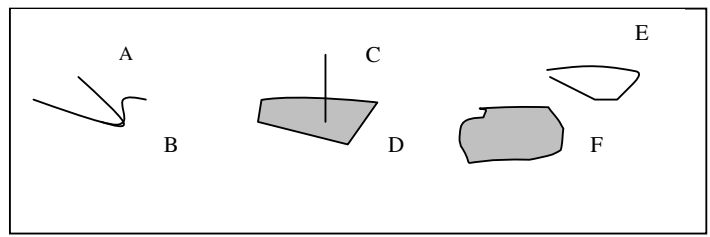

Figure 2.Examples of spatial scenes

- ✓ $SS_{AB}$= *(A, meet, B)*
- ✓ $SS_{CD}$= *(C, Intersect, D )*
- ✓ $SS_{EF}$= *(E, disjoin, F)*

#### 4.1.2 Topological distance between spatial scenes

In order to measure the topological distance between two scenes, we propose to use the conceptual neighborhood graph defined by [11]. The topological similarity measure between two scenes depends on the number of edges to traverse to go from one space to another predicate. Depending on the model of Egenhofer et al, the topological distance is equal to the number of edges separating the two spatial predicates in the conceptual neighborhood graph. [11] propose to decompose the conceptual neighborhood graph into three groups of topological relationships. In this context, the distance between two arcs in the same group (intra-group) is 2. However, the distance between two arcs belonging to two different groups is equal to 3. Except for the distance between the predicates (meet, meet and overlap) which has a processing cost equal owing diagram shows the distance in terms of number of edges separating different spatial predicates according to the TDD model. Figure 3 presents the Conceptual neighborhood network of the topological relationships according to the TDD model.

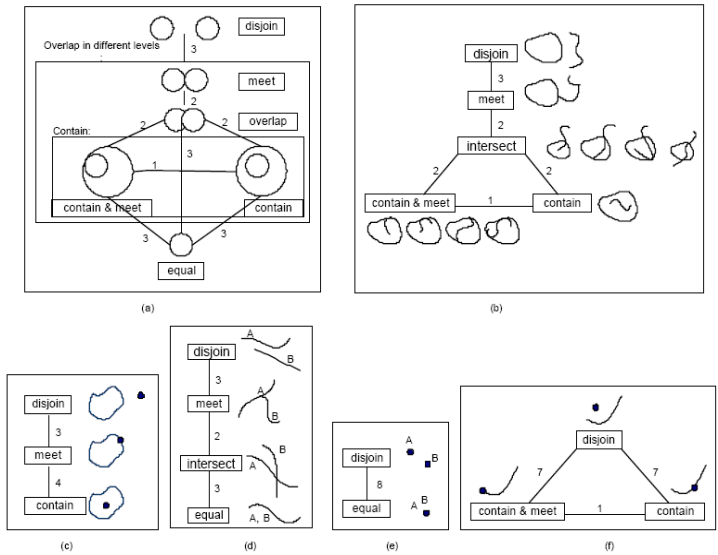

Figure 3. Conceptual neighborhood network of topological relationships: (a) two polygons; (b) a line and a polygon; (c) a point and a polygon; (d) two lines; (e) two points; (f) a point and a line [11]

**Definition:**

Given two spatial scenes $SS_{q1}$ and $SS_{q2}$ invoked respectively by the queries $q1$ and $q2$

$SS_{q1} = (Ob_a, Rt_{ab}, Ob_b)$ and $SS_{q2} = (Ob_c, Rt_{cd}, Ob_d)$

$$D_{top}(SS_1, SS_2) = MTC(Rt_{ab}, Rt_{cd}) \qquad (1)$$

With:

$D_{top}(SS_1, SS_2)$: The topological distance between the spatial scene $SS_1$ and the spatial scene $SS_2$

$MTC(Rt_{ab}, Rt_{cd})$: The minimum cost of transformation of the spatial relation $Rt_{ab}$ invoked by the query q1 to the the spatial relation $Rt_{cd}$ invoked in the query q2 according to the TDD model.

**Example:** Let the following three scenes $SS_{q1}$, $SS_{q2}$ and $SS_{q3}$ invoked respectively by *q1, q2 and q2 presented* in Section 2:

$SS_{q1}$: (zone1, meet, zone2)
$SS_{q2}$: (zone 2, intersect, zone 3)
$SS_{q3}$: (zone 4, disjoin, zone 5)

$D_{top}(SS_{q1}, SS_{q2})$=MTC (*meet*, *intersect*)=2
$D_{top}(SS_{q1}, SSq3)$=MTC (meet, disjoint)=3.

## 4.2. Distance between queries in term of the orientation

To measure the distance between two spatial queries in term of the orientation of the spatial objects, we propose to measure the distance between the spatial scenes invoked by each query. In the literature, nine types of directions are used namely: {north, northwest, west, southwest, south, southeast, east, northeast, and equality}. According to the TDD model, the cost of converting a direction into a close direction is equal to 2. The following diagram shows the cost of moving from one status to another.

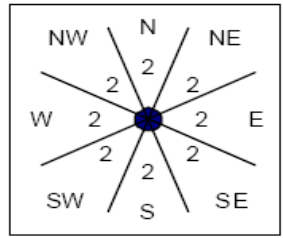

Figure 4. Graph of spatial directions and the costs of transformation according to the TDD model [11].

**Definition:**
Given two spatial MDX queries *q1 and q2*,
$SS_{q1}$, $SS_{q2}$ two spatial scenes invoked respectively by *q1* and *q2*
$ort_{SS1}$, $ort_{SS2}$: are respectively the orientation of the objects in the spatial scene $SS_1$ and the orientation of objects in the spatial scene $SS_2$.
$ort_{SS1}$, $ort_{SS2} \in$ {nord, nord-ouest, ouest, sud-ouest, sud, sud-est, est, nord-est, et l'égalité}

$$D_{dir}(q_1, q_2) = D_{dir}(SS_1, SS_2) = MTC(ort_{SS1}, ort_{SS2}) \quad (2)$$

With:

$D_{dir}(q_1, q_2)$: The distance in term of orientation between the query *q1* and the query *q2*.

$MTC(ort_{SS1}, ort_{SS2})$: The minimum cost of transformation of the orientation of objects in the spatial scene $SS_1$ to the orientation of objects in the spatial scène $SS_2$.

**Example:**

Given the queries *q1, q2* and *q3* presented in section x,

• The areas cited in the query *q1* are zone1 and zone 2
• The areas cited in the query *q2* are Zone 2 and Zone 3
• The areas cited in the query *q3* are zone 4 and zone 5

Suppose that on the topographic map (GIS map), the Zone 1, Zone 2, Zone 4 and Zone 5 are located as follows (Figure 5):

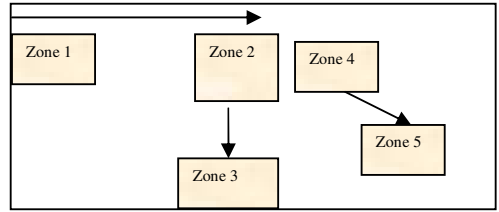

Figure 5. Schematic graph of the different locations of the areas in our example

*According to the orientation schema, we have:*

$ort_{SS1}$=orientation (zone 1, zone 2)= North
$ort_{SS2}$=orientation (zone 2, zone 3)=West
$ort_{SS3}$=orientation (zone 4, zone 5)= Northwest

So, we have:

$D_{dir}(q1, q2) = D_{dir}(ort_{SS1}, ort_{SS2}) = D_{dir}$ (Nord, Ouest)= 4
$D_{dir}(q1, q3) = D_{dir}(ort_{SS1}, ort_{SS3}) = D_{dir}$ (Nord, Nord-Ouest)= 2

We conclude that the query *q1* is more similar to the query *q3* regarding the directional distance

## 4.3. Distance between queries in term of the metric distance

To measure the metric distance between a pair of queries, we propose to measure the metric distance between scenes invoked by each query.

For this purpose, we propose to use the traditional model composed by four possible situations for the distances (equal, ready, average, below). The cost of transition from one situation to another is equal to 1. Figure 6 shows the various possible situations and the cost of transition from one situation to another basing on the TDD model.

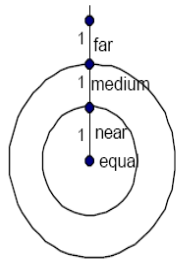

Figure 6. Graph of neighborhood and cost of transformation according to the TDD model (Li et al, 2006).

**Definition:**

Given two MDX queries *q1 and q2*

$SS_{q1}$, $SS_{q2}$ two spatial scenes invoked respectively by *q1* and *q2*

$OB_{SS1}=(ob_1^1, ob_1^2, ..., ob_1^i)$ with $ob_1^1, ob_1^2, ..., ob_1^i$: the spatial objects invoked in the spatial scene $SS_1$

$OB_{SS2}=(ob_2^1, ob_2^2, ..., ob_2^j)$ avec $ob_2^1, ob_2^2, ..., ob_2^j$: the spatial objects invoked in the spatial scene $SS_2$

*Let A=aij: the matrix used to measure the metric distance between the spatial objects of the query q1 and the spatial objects of the query q2, 1≤i≤n, 1≤j≤m and aij: the metric distance between the object i in the query q1 and the object j of the query q2*

The metric distance between the query *q1* and the query *q2* denoted $D_{met}$ *(q1, q2)* is computed as follows:

$$D_{met}(q1, q2) = \sum_{i=1}^{n}\sum_{j=1}^{m} a_{ij} \; ; 1 \leq i \leq n \text{ and } 1 \leq j \leq m \qquad (3)$$

**Example:**

Suppose that the disposal of Zone 1, Zone 2, Zone 3 and Zone 5 on the metric neighborhood graph is as follows:

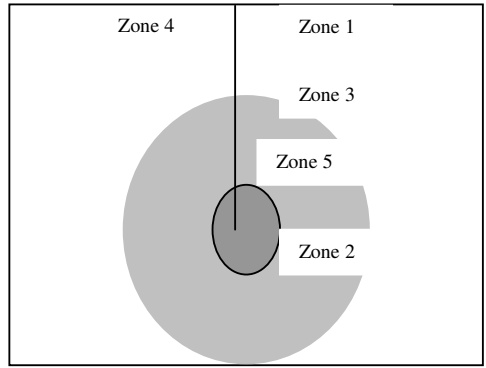

Figure 7. The metric Neighborhood graph in our example

The table corresponding to the matrix of the metric distance between q1 and q2 is as follows:

|        | Zone 1                           | Zone 2                            |
|--------|----------------------------------|-----------------------------------|
| Zone 2 | $D_{met}$ (zone1,zone2)= 3       | $D_{met}$ (zone2,zone2)= 1        |
| Zone 3 | $D_{met}$ (zone3,zone1)= 0       | $D_{met}$ (zone3, zone2)= 2       |

$D_{met}$ *(q1, q2)= $D_{met}$ (zone1,zone2)+ (zone3,zone1)+ $D_{met}$ (zone2,zone2) + $D_{met}$ (zone3, zone2) =* 3+1+0+2=6

The table corresponding to the matrix of the metric distance between *q1* and *q 3* is as follows:

|  | Zone 1 | Zone 2 |
|---|---|---|
| Zone 4 | $D_{met}$ (zone 4,zone1)=0 | $D_{met}$ (zone 4,zone2) =3 |
| Zone 5 | $D_{met}$ (zone 5,zone1) =2 | $D_{met}$ (zone5, zone2)=1 |

$D_{met}$ *(q1, q3)= $D_{met}$ (zone 4, zone1)+ (zone 4, zone2)+ $D_{met}$ (zone 5, zone1) + $D_{met}$ (zone5, zone2) = 0+3+2+1=6*

Thus the query *q1* has the same metric distance metric relative to the query *q2* and the query *q3*.

### 4.4. Spatial distance between MDX queries

**Definition:**
*Given q1, q2: two MDX queries. The spatial distance between q1 and q2 is computed as follows:*

$$D_{Spatial}(q1, q2) = D_{topo}(q1, q2) + D_{dir}(q1, q2) + D_{met}(q1, q2) \quad (4)$$

With:

$D_{topo}$ *(q1, q2):the topological distance between q1 and q2*
$D_{dir}$ *(q1, q2)*: *the distance in term of orientation between q1 and q2*
$D_{met}$ *(q1, q2)*: *the metric distance between q1 and q2*

**Example:**
Given *q1, q2* and *q3*

$D_{Spatial}$ *(q1, q2)= $D_{topo}$ (q1, q2) + $D_{dir}$ (q1, q2) + $D_{met}$ (q1, q2)= 6+2+4=12*
$D_{Spatial}$ *(q1, q3)= $D_{topo}$ (q1, q3) + $D_{dir}$ (q1, q3) + $D_{met}$ (q1, q3)= 6+3+2= 11*

### 4.5 Spatial similarity measure between MDX queries

The similarity is inversely proportional to the distance, the higher is the distance, the lower is the similarity and vice versa. Thus, we define the spatial similarity based on the spatial distance between queries as follows:

*Given two spatial MDX queries q1 and q2. The spatial similarity between q1 and q2 denoted $Sim_{spatial}$(q1, q2)is comptued as follows:*

$$Sim_{spatial}(q1, q2) = 1/(1+ D_{spatial}(q1, q2)) \quad (5)$$

## 5. PERFORMANCE EVALUATIONS

In this section, we present the experiments we have conducted to evaluate the performance of our proposed similarity measure. We generate 30 MDX queries produced by our own data generator. We adopt the Rubenstein and Goodenough [13] methodology for Human Relatedness Study: using 15 subjects for scoring 30 pairs of spatial MDX queries

In order to evaluate the performance of the proposed similarity measure we use the human evaluation technique through the Spearman's correlation coefficient. The Spearman's correlation coefficient is used to assess the degree of the closeness of the rankings of a set of data. The value of Spearman's correlation coefficient ranges from 1 to -1. A value of 1 indicates identical rankings, a value of -1 indicates exactly opposite rankings and a value of 0 indicates no correlation between the rankings. The other values of the coefficient indicate intermediate levels of correlation between these.

The 30 MDX queries have different degrees of semantic relatedness as assigned by the proposed similarity measure. The human subjects assign degrees of synonymy, on a scale from 0 to 4, We chose 10 pairs having a high degree of similarity according to our similarity measure (score between 3 and 4), 10 pairs having a weak degree of similarity scores (score between 1 and 3) and 10 pairs indicating an intermediate degree of relatedness (score between 0 and 1).

The experimental evaluation gave us a Spearman's correlation coefficient egal to 0,68 which reflects a high degré of similarity between the evaluated queries.

## 6. CONCLUSION

The research presented in this paper explores the role of the spatial similarity and proximity when applied to the development of SOLAP personalization techniques. The contribution of the paper can be summarized as follows. First, It proposes a user centric similarity measure between spatial MDX queries in the context of SOLAP manipulations. The spatial similarity measure addresses the basic spatial similarity component defined in the literature: (1) the spatial proximity (2) the distance in term of the orientation between the spatial objects and (3) the directional distance between the spatial MDX queries. To the best of our knowledge, our proposal is the first work proposing a similarity measure between spatial MDX queries for spatial personalization. Second, experimental evaluations are presented and validate the effeciency of our work. This proposal offers several research perspectives, in particular, we plan to extend the semantic component of the similarity between MDX queries basing on the ontologies. Moreover, the proposed similarity measure will be used for the development of a collaborative approach for the recommendation of spatial personalized MDX queries.